# Difference-in-Differences for Health Policy and Practice: A Review of Modern Methods

*Shuo Feng[1], Ishani Ganguli[2], Youjin Lee[1], John Poe[3], Andrew Ryan[4], Alyssa Bilinski[1,4*]*

[1] Department of Biostatistics, Brown University School of Public Health
2 Division of General Internal Medicine and Primary Care, Brigham and Women's Hospital, Harvard Medical School
3 Hessian LLC
4 Department of Health Services, Policy & Practice, Brown University School of Public Health
* Corresponding author: alyssa_bilinski@brown.edu

**ABSTRACT**

Difference-in-differences (DiD) is the most popular observational causal inference method in health policy, employed to evaluate the real-world impact of policies and programs. To estimate treatment effects, DiD relies on the "parallel trends assumption", that on average treatment and comparison groups would have had parallel trajectories in the absence of an intervention. Historically, DiD has been considered broadly applicable and straightforward to implement, but recent years have seen rapid advancements in DiD methods. This paper reviews and synthesizes these innovations for medical and health policy researchers. We focus on four topics: (1) assessing the parallel trends assumption in health policy contexts; (2) relaxing the parallel trends assumption when appropriate; (3) employing estimators to account for staggered treatment timing; and (4) conducting robust inference for analyses in which normal-based clustered standard errors are inappropriate. For each, we explain challenges and common pitfalls in traditional DiD and modern methods available to address these issues.

# I    INTRODUCTION

Difference-in-differences (DiD) is the most popular method for observational causal inference in health policy. DiD evaluates the impact of policies and programs using time-series data by comparing outcome trajectories between treatment and non-experimental comparison groups.[1] Its popularity partly stems from its flexibility and broad applicability: DiD allows for treatment effect estimation even when there exist no comparison groups that are exactly comparable to treatment groups. DiD instead relies on the assumption that treatment and comparison groups would have had parallel average outcome trajectories in the absence of a treatment effect (the counterfactual "parallel trends assumption"). Recent health policy applications of DiD include analyzing the impact of Medicaid expansion[2,3] and other insurance programs[4-7] on insurance coverage and health, same-sex marriage on mental health,[8,9] closures of automobile assembly plants on opioid overdoses,[10] sweetened beverage taxes on soda consumption,[11] and restrictions on sales of flavored tobacco products on youth smoking.[12]

Over the past several years, there have been rapid advancements in approaches for rigorously implementing and interpreting DiD.[13-17] In this paper, we review these recent innovations with the aim of synthesizing and presenting best practices for medical and health policy researchers. Although previous reviews have summarized these methods for an economics audience,[13-15] we adapt key points from prior reviews (e.g., Roth et al.[13]) and develop new material for the medical audience, extending prior tutorials in the health literature.[1,18,19] We cover common pitfalls related to: (1) evaluation of the parallel trends assumption[16,20-23] (2) adjustment for non-parallel trends,[24-28] (3) staggered treatment timing,[14,29-33] and (4) inference, detailing recent advancements that can address these common challenges in DiD applications.[34-38]

## II     DID ASSUMPTIONS AND ESTIMATION

The key identifying assumption in DiD is the "parallel trends assumption", which requires that on average, outcomes for treatment and comparison groups would have had parallel trajectories in the absence of the intervention under study. We also assume no anticipation (i.e., no effect in treatment groups prior to treatment) and no spillovers (e.g., no treatment effect in comparison groups).[13, 39] Combined, these assumptions allow us to estimate what would have occurred in treatment groups, absent intervention (dotted lines in Figure 1), and use this to estimate the average treatment effect on the treated (ATT).

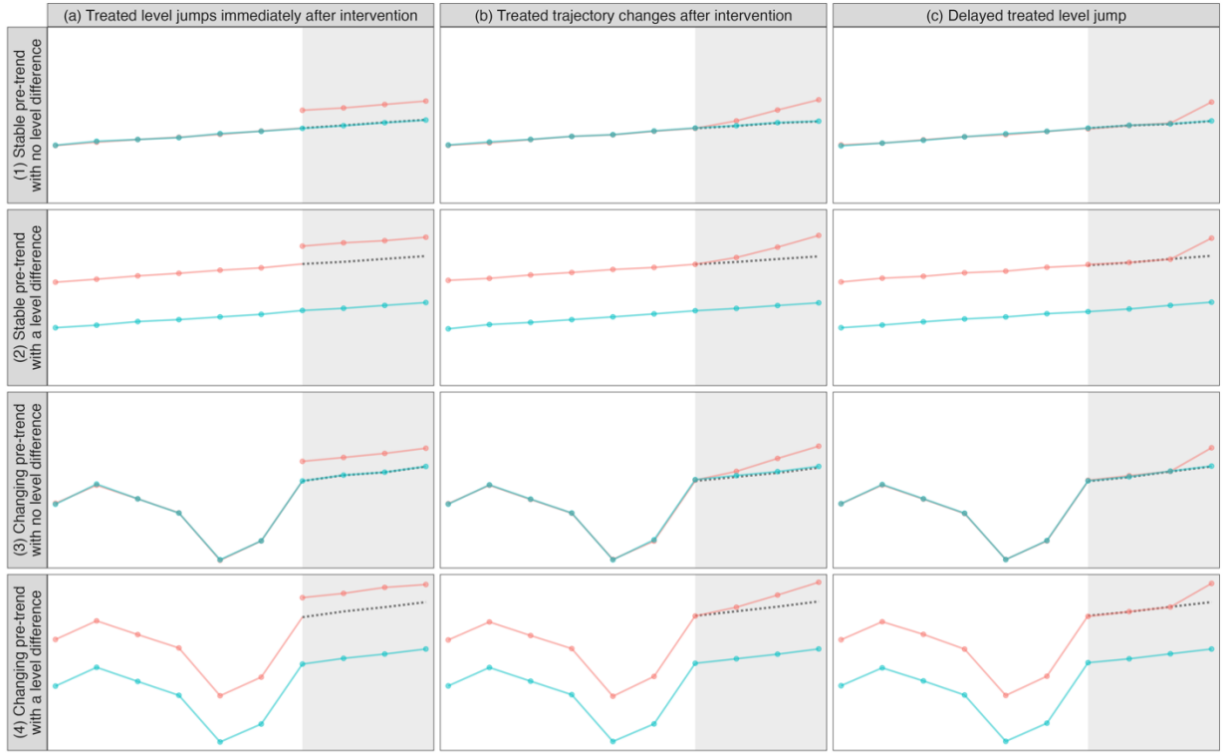

**Figure 1**: *Outcome trajectories under the parallel trends assumption.* Red and blue solid lines represent the observed outcome trajectories over time in treatment and comparison groups, respectively. The dotted black line indicates the untreated potential outcome in treatment groups assumed under DiD. The gray shaded area marks the post-intervention period.

The most popular specification used for DiD is the static two-way fixed effects (TWFE) model, estimated with ordinary least squares (OLS) regression:[13]

$$Y_{i,t} = \eta_i + \tau_t + \delta(Treated_i \times PostTreatment_t) + \epsilon_{i,t} \qquad \text{Eq. 1}$$

where $Y_{i,t}$ is the outcome for unit $i$ at time $t$, where $i = 1, \ldots, N$ (the number of units) and $t = 1, \ldots, T$ (the total number of time periods). $\eta_i$ is a unit fixed-effect for each unit $i$, $\tau_t$ is the a fixed-effect for each time period $t$, and $\epsilon_{i,t}$ is an idiosyncratic error for unit $i$ at time $t$. $Treated_i$ and $PostTreatment_t$ are binary variables indicating the treatment status for unit $i$ and time period $t$. In this setup, $\delta$ is the parameter of interest, which can correspond to the ATT in some simple cases (e.g., with unconditional parallel trends and non-staggered treatment timing). By specifying both unit- and time-level fixed effects, the TWFE model allows the outcome levels in the two groups to be different across units and to vary over time, provided counterfactual parallel trends. Later, we will discuss alternative specifications that address limitations of TWFE. Importantly, the terms "DiD" and "TWFE" are not interchangeable; TWFE is a simple estimation approach used in a subset of DiD analyses.

## III    EVALUATING THE PARALLEL TRENDS ASSUMPTION

### III.A   Pitfall 1: Not considering context or functional form

The parallel trends assumption is not directly testable because what would have happened in treatment groups without intervention cannot be directly observed. However, researchers often use evidence of parallel pre-intervention time series to help support its plausibility. In theory, parallel trends can be satisfied (1) with or without a level difference between treatment and comparison groups, and (2) with or without smooth pre-intervention trends (rows of Figure 1). Likewise, there can (3) be different patterns in treatment effects, including: (a) level changes immediately post-intervention, (b) trajectory changes immediately post-intervention, or (c) delayed level or trajectory changes (columns of Figure 1). However, DiD designs are often thought to be most compelling when level differences between groups are small before the intervention[16] and there is a sizable change in the treated outcome shortly after the intervention (e.g., Figure 1, column a). When there are increasing treatment effects (e.g., Figure 1, column b) as an intervention is phased in, researchers should investigate whether these might suggest continuation of pre-intervention trend or be driven by other post-intervention policies or changes. When treatment effects are delayed (e.g., Figure 1, column c), researchers should provide specific contextual evidence to support the observed effect timing (e.g., time needed for a drug to reach peak effectiveness) and again assess whether other contemporaneous events may have produced the observed effect.

These considerations speak to a larger challenge: the parallel trends assumption invites the question of why we believe treatment and comparison groups would have had similar counterfactual trends, even when their outcomes had different levels prior to the intervention.[16] Recent work has emphasized that researchers should draw on context-specific theory to provide evidence as to why treatment and comparison groups would be expected to trend in parallel in the absence of intervention.[16, 23] Ideally, researchers should be able explain why they would expect parallel trends prior to seeing the outcomes data.[13] Where context-specific theory is well-understood, researchers may also translate the parallel trends assumption into domain-specific conditions. For example, recent work has described the conditions required for modeling infectious disease outcomes using DiD.[23, 40] Such conditions can also help select functional form of the outcome, to which the parallel trends assumption is usually sensitive.[13, 28, 41] For instance, if the parallel trends assumption holds in the outcome variable's original levels, it will generally not hold in log-transformed levels (for other transformations, see[42, 43]), unless there was randomization of treatment, no time trends, or some combination thereof.[28]

### III.B   Pitfall 2: Using misleading statistical tests

Researchers frequently employ statistical tests to evaluate the evidence in favor of parallel pre-intervention trends. Traditionally, pre-trend tests have been based on a null hypothesis that there was no violation of parallel trends during the pre-intervention periods, e.g., they have asked whether the null effect was included in the 95% confidence interval of pre-intervention treatment effect estimates? However, as several papers have pointed out,[16, 20, 44, 45] these tests can be misleading because they may not have adequate statistical power to detect a violation, even if there are non-parallel trends between treatment and comparison groups.[16, 20, 44, 45] One alternative is to consider non-inferiority tests[20, 46] which specify the null to be a meaningful violation in the pre-treatment trend (instead of no violation) and reject only if there is strong evidence in the data that the violation in parallel pre-trends is small.[13, 20] A related approach is to report power of the pre-trends test against what the researcher specifies to be relevant plausible violations of parallel trends.[44] However, conducting statistical pre-trend tests and conditioning the analysis on surviving these tests may introduce bias in the estimated treatment effects even when there are no pre-existing differential trends between the two groups because the observed draws of the data which pass such tests are a selected sample from the true underlying data-generating process.[44]

Event study plots (e.g., Figure 2) are also a useful diagnostic for evaluating pre-intervention trends, which graphs the estimated treatment effects between treatment and comparison groups at each time point relative to the last pre-intervention time period (Eq. *2* in Appendix A). The pre-intervention points can be interpreted as placebo tests (i.e., should be small in magnitude), and the post-intervention points show the pattern of treatment effects. Though event study plots are not appropriate for the assessment of a no-violation null hypothesis using pre-treatment data owing to the power issue outlined above[20, 44] and may be less reliable with staggered treatment timing,[47] we encourage the use of them to visually assess whether there is any pre-existing trend between the two groups, to better understand the pattern of level or trajectory change in the treated outcome after the intervention (columns in Figure 2), and to examine bounds of confidence intervals to rule out unlikely pre-intervention trend patterns.

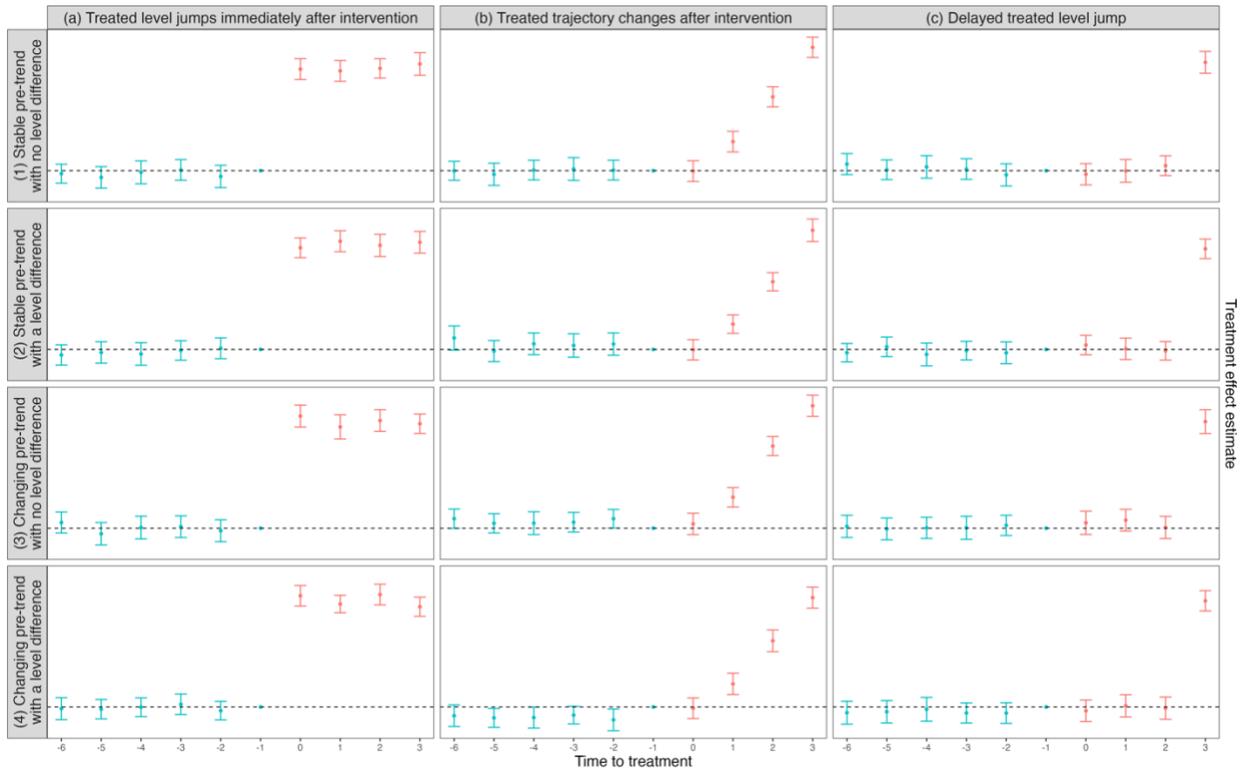

**Figure 2**: *Event study plots under different scenarios of parallel trends*. The event study plot in each panel corresponds to the outcome trajectories of the same panel in Figure 1. The dotted horizontal line represents 0, i.e., no treatment effect. The vertical error bars represent the 95% confidence intervals. Treatment effect estimates are generated according to Eq. *2* in Appendix A.

In summary, because the parallel trends assumption is not directly testable, researchers should justify why they expect treatment and comparison groups to share parallel trends even if pre-intervention levels differ. Conditions derived from context-specific theory can help reinforce the plausibility of this assumption. Researchers may also report results from non-inferiority tests and present event study plots to visualize any pre-existing trends and the pattern of post-treatment evolution in the treated outcome.

## IV   ADDRESSING NON-PARALLEL TRENDS

It is not uncommon for researchers to expect (or observe after initial evaluation of pre-intervention trends) that treatment and comparison groups had different trajectories, suggesting that the parallel trends assumption may not hold unconditionally. When this occurs, researchers often either (1) posit that the parallel trends assumption holds conditionally in sub-groups defined by baseline covariates and adjust for them,[48-52] or (2) derive bounds on the bias induced by violations of the parallel trends assumption.[53-55] In this section we will discuss two issues that can arise in this process: lacking a clear concept of confounding and adding regression coefficients directly to the TWFE specification.

### IV.A   Pitfall 3: Lacking a clear concept of confounding

For many traditional observational designs in which researchers evaluate the relationship between an outcome and a treatment of interest, any variable that affects both treatment status and the outcome can be a confounder, and failing to adjust for it may bias treatment effect estimates. By contrast, in DiD, covariates that produce only time-invariant level shifts between outcomes in the treatment and comparison groups are not confounders, and mechanistically, excluding them from a DiD specification does not bias ATT estimates (Table 1, row 1).

However, variables that affect the outcome trajectory can confound treatment effect estimates. We argue that researchers should focus on adjusting for baseline confounders for which they believe a "conditional parallel trends assumption" holds, i.e., trends are parallel within covariate-based subgroups measured at the beginning of the study (e.g., parallel trends among smokers and non-smokers). In practice, this leads to adjustment in two scenarios. First, in either cross-sectional or longitudinal/panel data, researchers should adjust for covariates that shift an outcome's slope or trajectory (Table 1, row 3). Second, when data is repeated cross-sectional rather than longitudinal/panel, researchers may need to account for differences in covariate distributions between treatment and comparison groups across time periods, even if covariates only affect outcome's level (Table 1, row 2).

We urge caution when adjusting for confounders that themselves vary over time (e.g., blood pressure) for two reasons (Table 1, row 4).[56] First, this invites a risk of adjusting for variables on the causal pathway if the confounder can itself be affected by treatment. Second, conditioning the parallel trends assumption on a time-varying variable makes it difficult to conceptualize a cohesive notion of underlying parallel trends – if trends are only parallel conditional on a time-varying variable, trends are not, in fact, believed to be parallel in any well-defined subgroup. As a result, in this case, other study designs may be more appropriate than DiD.

## IV.B  Pitfall 4: Adding regression coefficients directly to the TWFE specification

Mirroring methods used in other observational studies, many DiD studies add covariates directly into the TWFE specification (Eq. 1) to adjust for confounders.[13, 52] However, this approach does not appropriately adjust for confounding assuming parallel trends conditional on covariates. In fact, adding covariates to Eq. 1 may introduce bias if treatment effects change over time.[13]

Several techniques have been proposed in the literature to allow for consistent treatment effect estimation assuming conditional parallel trends. These fall into two main categories, which we denote trajectory modeling[24, 25, 49, 51, 57] and propensity score weighting.[13, 26]

Techniques in the first category involve modeling subgroup- or covariate-specific trends, using these to estimate subgroup effects, and combining these into an ATT.[24, 25, 49, 52] To do this, researchers first estimate covariate-specific trends in comparison group data. They then use these to estimate covariate-specific treatment effects, which can be combined to generate an overall ATT estimate.[13, 24] These methods perform well if covariate-specific trends are correctly specified. Matching is a special form of trajectory modeling, in which researchers define subgroup-specific trajectories based only on comparison units with similar baseline characteristics and compare these to corresponding treated units (i.e., DiD on a "matched" subsample of data).[49, 51, 57]

Alternatively, researchers can use inverse probability weighting (IPW) to recover the ATT assuming conditional parallel trends. This involves first fitting a propensity score model that predicts treatment status from baseline covariates, and using the estimated propensity scores to define weights that account for selection into treatment.[13, 26] This approach performs well if the propensity score model is correctly specified.

Doubly robust estimators combine trajectory modeling and propensity score estimation and perform well if either the trajectory model or the propensity score model is correctly specified.[25]

Other innovations in confounder adjustment methods include Bayesian approaches for trajectory modeling,[58, 59] machine learning to estimate weights for IPW and doubly-robust estimators,[60-62] and IPW approaches driven by latent variables or marginal structure models to facilitate a better understanding of selection into treatment.[62-64]

Last, beyond or in lieu of adjusting for specific covariates, researchers can use sensitivity analyses to quantify the extent to which conclusions are sensitive to violations of parallel trends. For example, Rambachan and Roth[54] proposed an approach to bound how large the post-treatment violation of parallel trends must be relative to the highest pre-treatment violation to meaningfully shift effect estimates and developed methods to construct valid confidence intervals for the treatment effect under the imposed restrictions, even if the parallel trends assumption is violated. Keele et al.[65] likewise outlined a sensitivity analysis based on matching that allows researchers to assess how strong the unobserved confounders must be to alter or reverse their conclusions. Third, Ye et al.[55] proposed bounding bias based on the most and least conservative comparison groups. These approaches allow researchers to evaluate and communicate the robustness of conclusions across reasonable assumptions even lacking data on potential confounders.

In summary, when faced with violations of parallel trends, researchers might carefully consider potential confounders and use trajectory modeling, propensity score weighting, or combined doubly robust estimators to account for them. Sensitivity analyses can also help quantify robustness of key conclusions to violations of parallel trends.

# V DID WITH STAGGERED TREATMENT TIMING

## V.A Pitfall 5: Using static TWFE with staggered treatment timing

Researchers frequently apply DiD to settings in which units initiated treatment at different calendar times. When this is the case, conducting DiD with a static TWFE specification (Eq. 1) can produce a biased ATT estimate unless treatment effects are homogeneous over time and across all treatment adoption cohorts (Table 2, row 1).[13] This occurs because the treatment effect estimate produced by TWFE is a weighted average of all possible 2 × 2 comparisons over all groups and treatment times, with early treated units used as comparison units after treatment.[32] When the treatment effect is changing over time, this can even flip the sign of the treatment effect.[13, 30, 32] Furthermore, 2x2 estimates are combined into a treatment effect estimate based on weights designed to minimize variance. When treatment effects differ across adoption cohorts, this treatment effect estimand may no longer reflect the overall ATT of interest.

One simple extension to the static TWFE specification (Eq. *1*), which partially addresses these problems, is the dynamic TWFE specification (Eq. 3 in Appendix B). In dynamic TWFE, the single treatment status indicator is replaced by multiple indicators, providing a treatment effect for each post-intervention time period relative to treatment initiation (Table 2, row 2). The dynamic TWFE specification yields a sensible estimand for the ATT if heterogeneity only exists in the number of time periods since treatment, but it assumes that average treatment effects are the same across all cohorts regardless of adoption time.[31, 66] For example, dynamic TWFE allows treatment effects to be different in year 2 compared to year 3 following treatment adoption, but the year 2 effect is assumed the same across all adoption cohorts.[13] This may not be plausible if early adopters had a different experience of the policy (e.g., a pilot phase) or if there was selection bias into treatment.

Nevertheless, there are other methods that allow for treatment effect heterogeneity both across cohorts and over time. In stacked regression, researchers create a new dataset by stacking each treated cohort with the group of comparison units that were not-yet-treated over a researcher-selected time horizon (e.g., four time periods before and two time periods after treated group received the treatment).[67] This approach allows both time and cohort heterogeneity, but it does not provide users with the flexibility to specify weights to aggregate treatment effects into the ATT; instead, the contribution to the treatment effect in a stacked regression relies on the variance of the treatment effect estimate from each stacked dataset, even though it is possible for researchers to reweigh as they see fit.[68]

For further flexibility, researchers can generate a treatment effect estimate for each cohort at each post-intervention time (often called a "cohort-time" or "group-time" treatment effect in the literature[30, 31]) and aggregate these as desired. There is growing consensus that researchers should default to these methods, which include estimators proposed by Callaway and Sant'Anna, Sun and Abraham, and Borusyak, Jaravel, and Spiess (Table 2, rows 3 and 4).[30,31,66] They all construct cohort-time treatment effect estimates, but differ in the choice of comparison groups and the pre-intervention time period.[13, 69, 70] For example, the Callaway and Sant'Anna[30] estimator uses only the last pre-treatment period in the comparison groups, which invokes a weaker parallel trends assumption but may have correspondingly lower power.[70]

Overall, these alternatives provide well-defined causal parameters with transparent weights over 2 × 2 comparisons and avoid using treated units as comparison groups.[13]

# VI DID WITH ROBUST INFERENCE

## VI.A Pitfall 6: Using normal-based clustered standard errors with insufficient numbers of treatment and comparison groups

In DiD studies, treatment generally occurs at an aggregated level (e.g., state), but data may be collected either at the same aggregated level or a lower level of granularity. For example, we may have patient-level data, while the intervention was implemented at the hospital level, or we may have county-level data while the policy was implemented by state. In all scenarios, it is recommended to estimate the standard errors clustered at the level of treatment assignment.[13,71,72] Normal-based clustered robust inference (e.g., *cluster robust* in Stata) is one of the most popular methods for clustering standard errors. However, it requires a large number of both treated and untreated clusters (e.g., at least 25 or 30 clusters in each class), which may not exist in many DiD applications.[13,73] If there are only few clusters in the data, normal-based clustered standard errors are typically anti-conservative,[73] leading to inappropriately small p-values and inflated Type I error (Table 3, row 1).

Several alternative methods have been proposed to conduct inference with a small number of clusters. One popular approach is the wild cluster bootstrap (or wild score bootstrap for generalized linear models).[37] Simulations have suggested that this method can perform well with as few as five clusters, although its estimates may be conservative when the proportion of treated units is small (Table 3, row 2).[35,73,74] However, while simulation evidence shows strong performance, this method invokes an assumption of a balanced proportion of treated units among clusters, which may not be plausible in many DiD applications where the treatment is assigned at the cluster level (see Canay et al.[75]).

Other methods rely on different assumptions (Table 3, rows 3 and 4). For example, when there are many untreated clusters available but only few treated clusters, Conley and Taber proposed an approach to estimate the distribution of errors in treated clusters from those in untreated clusters.[36] Alternatively, with many time periods and the proportion of treated units in each cluster varying little over time, conformal inference can be a robust approach.[38]

Taken together, researchers should be wary of normal-based clustered standard errors if there are insufficient numbers of treatment or comparison groups and explore other alternatives to robust inference as needed.

## VII    CONCLUSION

This paper identifies the key pitfalls in DiD studies and compiles solutions that have been proposed in recent literature. Researchers should start with evaluating parallel trends based on context, visualizing data in plots, using statistical tools with adequate power, and selecting functional form guided by domain-specific knowledge. There are ways to relax the parallel trends assumption, such as by establishing parallel trends in subgroups of the data. We emphasize that this should be done by first considering the type of expected confounding and adopting appropriate machinery to properly adjust for the covariates if needed. When treated units initiate treatment at different times, we recommend researchers to default to group-time treatment effect estimation methods that perform well with heterogeneous treatment effects across time and cohorts. Finally, normal-based cluster inference requires many both treated and untreated clusters. When this is not the case, researchers should opt for alternatives that have weaker assumptions. Overall, we are optimistic that these recent innovations can strengthen DiD practice and associated policy recommendations.

ACKNOWLEDGEMENTS

The authors gratefully acknowledge feedback from Jason Buxbaum, Jonathan Roth, and Pedro Sant'Anna. This work was supported by the Centers for Disease Control and Prevention through the Council of State and Territorial Epidemiologists (NU38OT000297-02, S.F. and A.B.), the National Institute of Diabetes and Digestive and Kidney Disease (R01DK136515, Y.L.), and the National Institute on Aging (K23AG068240, I.G.).

| Label | Narrative | | How does parallel trends assumption hold? | Longitudinal (panel) vs. cohort data | Baseline covariate vs. repeated measures | Covariate effect | Recommended adjustment method(s) | Impact of naïve adjustment |
|---|---|---|---|---|---|---|---|---|
| | *Description* | *Example* | *Unconditionally, conditionally, or else?* | • **Longitudinal:** *The same group of individuals is followed over time.* <br> • **Cohort:** *Different individuals are sampled for treatment and comparison groups over time.* | • **Baseline:** *The covariate takes one value per individual, and can be measured at baseline.* <br> • **Repeated measures:** *The covariate can take different values per individual.* | *Does the covariate shift the outcome's level or its trajectory?* | *What adjustment (if any) is necessary to produce an unbiased treatment effect estimate?* | *What are the impacts of naïve regression adjustment on the treatment effect estimate?* |
| Longitudinal level differences | The same group of individuals is followed over time in treatment and comparison groups. Baseline covariates are measured once and only shift the outcome's level. | A study follows the same individuals in treatment and comparison groups over time. Covariates are the individual's inherent characteristics (e.g., gender, birth year) that only shift the outcome's level. | Unconditionally | Longitudinal | Baseline | Level | Not needed | No change in treatment effect (no bias). |
| Cohort level differences | Different individuals are sampled for treatment and comparison groups over time. Baseline covariates are measured once and only shift the outcome's level. | A survey is conducted every 6 months. Population-level characteristics (e.g., proportion of individuals in each age group) vary between rounds of the survey. However, at the respondent level, covariates values do not vary over time. | Conditional on the baseline covariates | Cohort | Baseline | Level | • Trajectory modeling* <br> ◦ Matching <br> • Inverse probability weighting <br> • Doubly-robust estimators | Adjustment changes treatment effect but will remain biased if changes in covariate values differ between treatment and comparison groups. |
| Trajectory differences | Baseline covariates take only one value per individual but shift the outcome's trajectory. | There are different outcome trajectories among rural and urban respondents, absent treatment, and the comparison group has a greater proportion of urban enrollees. | Conditional on the baseline covariates | Either | Baseline | Trajectory | • Trajectory modeling* <br> ◦ Matching <br> • Inverse probability weighting <br> • Doubly-robust estimators | • Longitudinal data: no change in treatment effect and remain biased. <br> • Cohort data: adjustment changes treatment effect with risk of introducing additional bias if changes in covariate values differ between treatment and comparison groups. |
| Repeated measures | Covariates are measured at multiple times. Their values vary across different measurements. | A clinical trial where the covariates include repeated measurements of the patient's vital signs such as blood pressure. These values affect the outcome. | Not recommended: it is difficult to envision how parallel trends would hold with repeated measures | Either | Repeated measures | Either | DiD may not be a reliable method due to: <br> • risk of reverse causality <br> • lack of cohesive notion of parallel trends | • Adjustment changes treatment effect but will remain biased if covariate values differ between treatment and comparison groups. <br> • There is also risk of introducing additional bias if treatment effects are not saturated (i.e., not estimating a treatment effect in each post-intervention period). |

* also referred to as regression adjustment in the literature.

**Table 1**: *Summary of DiD confounding*

| Label | Narrative | | Is static TWFE biased? | Heterogenous treatment effect in time since treatment | Heterogenous treatment effect in cohorts | Parallel trends assumption | Comparison group | Recommended adjustment method(s) |
|---|---|---|---|---|---|---|---|---|
| | *Description* | *Example* | *Yes/No* | *Are treatment effects different for different lengths of time since treatment?* | *Are treatment effects different across different cohorts?* | *Are parallel trends required on all time periods or post-treatment periods only\** | *What is used as the comparison group?* | *What adjustment (if any) is necessary to produce unbiased treatment effect estimate?* |
| Homogeneous treatment effects across unit and time | Treatment effects are the same regardless of how long the treated unit has received the treatment and across cohorts. The parallel trends assumption is imposed on all cohort averages and all time periods. | The effect of mask mandate is the same regardless of when or how long the county has implemented the policy. The average outcomes for all counties would have evolved in parallel absent the intervention. | No | No | No | All time periods | Weighted average of both not-yet-treated and any never-treated groups | Not needed |
| Homogeneous treatment effects over time only | Treatment effects can differ according to length of time since treatment, but must be the same across cohorts. The parallel trends assumption is imposed on all units and time periods. | The effect of mask mandate is stronger in the counties that have implemented the policy for a longer time, but it is the same for all counties at 1 month post-intervention despite the timing of adoption. The average outcomes for all counties would have evolved in parallel absent the intervention. | Yes | Yes | No | All time periods | Weighted average of both not-yet-treated and any never-treated groups | • Dynamic TWFE |
| Heterogenous treatment effects | Treatment effects can vary across different cohorts and over time. Still, the parallel trends assumption is imposed on all units and all pre-treatment periods. | The effect of mask mandate can depend on both the timing of adoption or how long the counties have implemented the policy. The average outcomes for all counties would have evolved in parallel absent the intervention. | Yes | Yes | Yes | Varies | Varies | • Imputation method by Borusyak, Jaravel and Spiess[66] (comparison group: average of all never-treated and not-yet treated)<br>• Stacked regression by Cengiz, et al.[67] (comparison group: never-treated or not-yet-treated)<br>• Sun & Abraham[31] (comparison group: never-treated or last-treated)<br>• Callaway and Sant'Anna[30] (comparison group: never-treated) |
| Heterogenous treatment effects with parallel trends assumed beginning at last-pre-intervention period | Treatment effects can vary across different cohorts and over time. The parallel trends assumption is only imposed on post-treatment periods and the last pre-treatment period. | The effect of mask mandate can depend on both the timing of adoption or how long the counties have implemented the policy. The parallel trends assumption can be imposed either on not-yet-treated units, never-treated units, or a combination of both. | Yes | Yes | Yes | Post-treatment periods and the last pre-treatment period | Pick not-yet-treated, never-treated, or a combination | • Callaway and Sant'Anna[30] (comparison group: not-yet-treated) |

\* The last pre-intervention period is often required for parallel trends if the entire pre-intervention periods are not required

**Table 2:** *Impacts of heterogeneous treatment effects across staggered treatment timings*

| Label | Examples of popular Implementation | Data requirements<br>*Consider:*<br>• *Number of clusters*<br>• *Size of each cluster*<br>• *Number of time periods* | Heterogeneous treatment effects<br>*Do we allow different clusters to have different treatment effects?* | Other assumptions<br>*What are some other assumptions required by the approach?* |
|---|---|---|---|---|
| **Normal-based clustered standard error** | • Sandwich estimators for standard errors:<br>  • `vcovCL(cluster = clustvar)` in R<br>  • `vce(cluster clustvar)` in Stata | • Large number of treated and untreated clusters | Yes | • Cluster-level errors follow an asymptotically normal distribution |
| **Bootstrap-based** | • Wild cluster bootstrap (OLS)[37]<br>• Wild score bootstrap (GLM)[37] | • Can work well with as few as 5 large clusters<br>• Clusters have balanced proportions of treated units<br>• The proportion of treated clusters is not small | Yes | • Estimates will be overly conservative if the treatment proportion of treated clusters is too small |
| **Permutation-based** | • Hageman (2022)[76]: bound on the maximal relative heterogeneity between clusters | | • Untreated potential outcomes cannot have any cluster-specific heterogeneity | • As cluster size grows large, any single untreated cluster could be used to infer the untreated potential outcomes for treated clusters |
| | • Fisher randomization test (FRT)[77] | | Yes | • Assumes random treatment assignment<br>• Inference only available under the sharp null of no treatment effects for all units<br>• Exchangeability between treated and untreated clusters |
| **Other methods** | • Conley and Taber (2011)[36]: learn the distribution of errors in treated clusters using that distribution from untreated clusters | • Large number of untreated clusters | • Yes, but heterogeneity in treatment effects often violates the assumption that errors in treated clusters come from the same distribution as those in untreated clusters | • Treated and untreated clusters have the same distribution of errors |
| | • Chernozhukov, Wüthrich and Zhu (2021)[38]: conformal inference | • Large number of time periods | • Yes, provided that the proportion of treated units in each cluster remains the same over time | • Parallel trends assumption needs to hold in many pre- and post-treatment time periods |

**Table 3:** *Summary of DiD inference methods*

# SUPPLEMENTARY MATERIAL

## A  EVENT STUDY

Events study plots build on the dynamic TWFE specification which extends the static TWFE by considering the effect of each cohort relative to the time of intervention:

$$Y_{i,t} = \eta_i + \tau_t + \sum_{r \neq -1} \beta_r \mathbb{I}(R_{i,t} = r) + \epsilon_{i,t}, \qquad \text{Eq. 2}$$

where $R_{i,t}$ denotes the time to intervention for unit $i$ at time period $t$, and $\beta_r$ is the coefficient on each time period relative to the time of intervention (e.g., $\hat{\beta}_r$ for r < 0 are the estimated $\beta_r$ measures of effects prior to the intervention), which is plotted in the y-axis of event study plots as displayed in Figure 2. In practice, $\hat{\beta}_{-1}$ is often normalized to 0 by convention.

## B  DYNAMIC TWFE SPECIFICATION

In a dynamic TWFE specification, we regress the outcome variable on the unit- and time-level fixed effects and dummies indicating the time relative to intervention that estimates a treatment effect at each post-intervention time period:

$$Y_{i,t} = \eta_i + \tau_t + \sum_{r \geq 0} \beta_r \mathbb{I}(R_{i,t} = r) + \epsilon_{i,t}, \qquad \text{Eq. 3}$$